\documentclass[aps,prd,twocolumn,groupedaddress]{revtex4-1}

\usepackage{graphicx}
\usepackage{subfigure}
\usepackage{amsmath} 

\usepackage[colorlinks=true,citecolor=blue,urlcolor=magenta,breaklinks]{hyperref}
\usepackage{natbib}
\usepackage[T1]{fontenc}
\usepackage{times}
\begin{document}

%\preprint{}
%\title{New neutrino physics and the altered shapes of solar  neutrino spectra}
%\author{Il\'idio Lopes\altaffilmark{1,2}}
%\altaffiltext{1}{Centro Multidisciplinar de Astrof\'{\i}sica, Instituto Superior T\'ecnico, 
%Universidade de Lisboa , Av. Rovisco Pais, 1049-001 Lisboa, Portugal} 
%\altaffiltext{2}{E-mail:ilidio.lopes@tecnico.ulisboa.pt} 
%
\title{Quadrupole stellar oscillations: The impact of gravitational waves from the Galactic Center}
\author{Il\'idio Lopes}
\email[]{ilidio.lopes@tecnico.ulisboa.pt}
\affiliation{Centro Multidisciplinar de Astrof\'{\i}sica, Instituto Superior T\'ecnico, 
Universidade de Lisboa , Avenida Rovisco Pais, 1049-001 Lisboa, Portugal}
%
%
%\date[Published in Phys. Rev. D 88, 045006, ]{7 August 2013}
% 

\begin{abstract}
Acoustic quadrupole modes of sunlike stars vibrate when perturbed by a passing gravitational wave generated somewhere in the Universe. Here, we compute the imprint of the gravitational waves on the acoustic spectrum of these stars for gravitational events occurring near the supermassive black hole located at the center of the Milky Way.  
We found that in most cases the impact of gravitational waves in low-order quadrupole modes is not above the current observational threshold of detectability, although this should be in the reach of the next generation of near infrared observatories and asteroseismology  satellite missions. Equally, we found that it is possible to follow the end phase of the coalescence of binaries with large chirp masses, as these phenomena have a unique imprint in the spectra of sunlike stars affecting sequentially several low-order quadrupole modes.  Moreover, we discuss the different imprints on the acoustic spectra of the different types of binary systems constituted either by two white dwarfs, two neutron stars, two black holes  or a compact star and a massive black hole.
\end{abstract}

%\keywords{}
%\keywords{Neutrinos -- Sun:evolution --Sun:interior -- Stars: evolution --Stars:interiors}

\maketitle


%%%%%%%%%%%%%%%%%%%%%%%%%%%%%%%%%%%%%%%%%%%%%%%%%%%%%%%%%%%%%%%%%%%%%%%%%%%%%%%%%%%%%%%%%%%%%%%% 
%%%%%%%%%%%%%%%%%%%%%%%%%%%%%%%%%%%%%%%%%%%%%%%%%%%%%%%%%%%%%%%%%%%%%%%%%%%%%
\section{Introduction}
%%%%%%%%%%%%%%%%%%%%%%%%%%%%%%%%%%%%%%%%%%%%%%%%%%%%%%%%%%%%%%%%%%%%%%%%%%%%%
The Milky Way harbors a supermassive black hole at its core.
Its location coincides with a well-known compact radio source, Sagittarius A$^*$~\citep{2008Natur.455...78D}. 
With a diameter of 0.3 AU this source is surrounded by one of the most dense stellar populations of the Galaxy. 
The most compelling proof of the existence of a black hole in Sagittarius A$^*$ is the continuous observation of the nearest stars orbiting 
in a very fast Keplerian motion which is only possible with a very concentrated massive object located in the Galactic Center~\citep{2005ApJ...628..246E}.
For example, a star of 15 $M_\odot$, known as S2, spins around the central object in an  elliptical orbit with a period of 15.2 yr 
and a  pericenter of 120 A.U. (smallest  distance from the central object).  
From the motion of S2,~\citet{2008ApJ...689.1044G} made the first 
estimation of the mass of the central black hole. These authors have determined this mass to be  $4.1\;10^6\;M_\odot$. 
More recently, from the measurement of the proper motions of several thousand stars within  one parsec from the central black hole,~\citet{2009A&A...502...91S} have estimated simultaneously the black hole's mass at $3.6^{+0.2}_{-0.4}\;10^{6}\; M_\odot $ and an additional distributed  mass of  $1.0\pm 0.5 \;10^{6}\; M_\odot $.
The latter mass term is due to a local population with a few tens of million of stars. The population consists of metal-rich, M, K, and G old giant stars, main-sequence B stars~\citep{2009ApJ...694...46D} and compact stars (or stellar remnants).

%%%
\smallskip
Equally, like the Milky Way many other massive galaxies are known to have a core made of a 
supermassive black hole surrounded by a concentrated and  dense stellar environment.
Throughout the lifetime of a galaxy, the central black hole grows by capturing many of the 
neighboring stars and clouds of molecular gas. 
A complex network of gravitational interactions between stars 
occurs continuously in the galactic core.
During these stellar encounters, a countless number of binaries form between 
neighboring pairs of white dwarfs (WD-WD), neutron stars (NS-NS), and  stellar black holes (BH-BH).
The recent discovery of gravitational radiation from the merger of two stellar black-hole binaries, 
ensures that the last type of binaries must be quite common in the Galaxy~\citep{2016PhRvL.116f1102A,2016PhRvL.116x1103A}. 
Another type of binary that is created with regularity, corresponds to the ones that are formed between the supermassive black hole and nearby compact stars.
In binaries like these for which the lighter star is 4 orders of magnitude less massive than the companion,  the associated gravitational event is classified as an extreme mass ratio inspiral (EMRI). 
As described above, stellar (compact) binaries and EMRIs should form in large numbers 
in the core of the Milky Way.

%%% 
\smallskip 
Nonradial oscillations have been discovered in many stars in the Milky Way
by the COROT~\citep{2006ESASP1306...33B} mission and   {\sc Kepler} 's main and 
extended missions~\citep{2010ApJ...713L.160G,2014PASP..126..398H}.
More than 18 000 main-sequence, subgiants and red giant stars   
have been shown to have oscillation spectra with identical properties.  
The combined spectra of these stars spans the frequency range from
$10^{-7}\,{\rm Hz}$ to $10^{-2}\,{\rm Hz}$~\citep{2011MNRAS.414.2594H,2014ApJS..210....1C}. 
The excitation and damping of these oscillations is attributed to the turbulent motions of convection in the external layers of these stars. These physical processes that excite stellar oscillations
are identical to the ones found for the Sun. For that reason such oscillations are known as sunlike oscillations
and the stars as sunlike stars. These stars are found in many regions of the Milky Way. The  {\sc Kepler}  mission alone has measured oscillations in stars located at distances up to 15 kiloparsecs. In many cases the sunlike stars belong to stellar clusters found in the direction of the Galactic Center, several of which are located above the galactic disk or in the bulge. Furthermore,  by making high precision observations in many new directions of the Galaxy, the future {\sc PLATO}  satellite~\citep{2013EGUGA..15.2581R} should be able to increase significantly the quantity and quality of sunlike stars discovered in the Milky Way.

\smallskip 
This work focuses on the study of the impact of  gravitational waves emitted by compact binaries (with special focus on EMRIs  events) in the spectra of sunlike stars. The motivation for the project comes from the fact that the core of the Milky Way is densely populated by compact stars, stellar binaries, and possibly multistellar systems, 
all of which are being attracted by the supermassive black hole. The gravitational radiation emitted by these binaries will excite some of the oscillations of the sunlike stars located at relatively short distances from the Galactic Center.

%%%
\smallskip 
Specifically, we are interested in studying the impact of the gravitational waves emitted 
during the binary contraction,  a phase known as the {\it inspiral phase}.
For our convenience this phase is  split into the following two stages:(1) the inspiral phase begins when the two stars are far-apart rotating in near-perfect circular orbits ({\it monochromatic emission})  and (2) finishes at the end of the orbits contraction ({\it chirp emission}), just before the merger of the two stars. There is a post-merger emission known as the ring-down phase, where the shape distortion of the nearly formed object is also dissipated as gravitational radiation, but this stage is not discussed in this work. 

\smallskip 
All the  different types of binary systems in the {\it inspiral phase} emit gravitational waves with a characteristic waveform known as "chirp" whose amplitude and frequency increases with time until the coalescence. Even though the observation of strongest events are expected to be more sporadic, the detection of such phenomena would produce very unique and interesting results that can be used to test General Relativity. For instance, in the case of EMRIs, these waves can be used to probe the gravitational field of the central black hole. In addition, as these waves travel large distances through space, equally these can be used to test the wave properties of gravitational radiation. 
Theoretical models predict gravitational chirps with strains on Earth of the order of  $10^{-17}$ -- $10^{-22}$, and frequencies varying from $10^{-4}$ to $10^{-1}$ Hz. Most of these events are expected to be detected by experiments like eLISA~\citep{2012CQGra..29l4016A, 2013LRR....16....7G,Moore:wk}.         
Nevertheless, the strain of such gravitational events will be much larger for stars located nearby these binaries.

%%%%% 
\smallskip 
Oscillating stars are natural receptors of gravitational waves. Like any resonant sphere, they have an isotropic sensitivity to gravitational radiation --  able to absorb gravitational waves from any direction of the sky. 
For a given frequency of acoustic oscillations, stars have sensitivity much larger than current resonant mass detectors, as their integrated scattering cross section and Q quality factor for gravitational waves is 20 and 2 orders of magnitude larger than the usual resonant mass detectors~\citep{1996PhRvD..54.1264M,2006CQGra..23S.239A,2007PhRvD..75b2002G,2015ApJ...807..135L}, respectively.  
For that reason in the rest of this article we will refer to these as {\it star detectors}.

 %%%%%% 
\smallskip  
The first studies of the absorption of gravitational waves by astrophysical objects like Earth, Moon, planets, and stars were developed  by~\citet[][]{1969ApJ...156..529D,1980ApJ...241..475Z,1997A&A...321.1024K}, among others. \citet{1984ApJ...286..387B} were the first to use helioseismology to constrain the amplitude of gravitational waves. Recently~\citet{2011ApJ...729..137S} 
and~\citet{LopesSilkGW2014,2015ApJ...807..135L} have updated these calculations. By using the unseen solar gravity modes, \citet{2014ApJ...784...88S} determined the maximum amplitude of the strain for the stochastic gravitational wave background~\citep{2010MNRAS.408.1742S,2012AN....333..978S}. By focusing on the study of the Sun as a natural gravitational wave detector~\citet{LopesSilkGW2014,2015ApJ...807..135L}   have shown the potential of asteroseismology for gravitational wave searches.
\citet{2014MNRAS.445L..74M} ) were the first to propose that stars near massive black-hole binaries could be efficient resonant GW detectors.
Moreover, these authors estimated that the gravitational radiation is absorbed by stars and have shown how such spectra could be observed by second generation space-born gravitational wave detectors. 

%%%%%%  Ilidio
\smallskip  
In this article, we study the impact that the gravitational wave emissions coming from binaries located in the core of Milky Way have on the low-order quadrupole acoustic modes of nearby sunlike stars. 
We found that for the more massive binaries (including EMRIs), it will be possible to follow the end of the inspiral phase, during which  a large emission  of gravitational radiation is expected in the last stage of the binary contraction (chirp emission), before the coalescence of the two black-holes.
We show for the first time that  the chirp waveform of the gravitational wave has an unique impact on the acoustic spectrum of the sunlike stars by
ringing up one quadruple mode after another.
Some gravitational chirp events could excite several modes in the same star. Although these gravitational emissions have relatively large strains, their impact on acoustic modes of nearby sunlike stars leads to relatively small amplitude variations, which we can expect to be well within reach of the next generation of near infrared observatories.

 %%%%%% 
\smallskip 
In this study the focus is on the impact of  gravitational radiation of low-order quadrupole modes, since recent developments in analyzing asteroseismology data and in our understanding of the theory of stellar pulsations have been quite successful in predicting the properties of acoustic modes.  Nevertheless, there is a large research potential in  studying the impact of gravitational waves in quadrupole gravity modes and mixed modes of sunlike stars in the main sequence or  in the red giant branch. This is particularly so as there is a significant amount of data available from recent  asteroseismology surveys. In particular, the impact of 
gravitational waves in gravity modes, more precisely in the  
Sun, was recently computed by~\citet{2011ApJ...729..137S}. 
They found  the velocity amplitude to  be of the order  of 
$10^{-5}$ -- $10^{-3}$ mm s$^{-1}$   for the Sun, but  
has the potential to be more significant for other stars.
Moreover, the same authors have shown that gravity modes 
can be used as an independent method to put an upper bound to 
the stochastic background of gravitational radiation~\citep{2014ApJ...784...88S}.

\smallskip  
In Sec.~\ref{sec-GWS}, we discuss a basic description of the formation and evolution of a binary system and the generation of gravitational waves. In Sec.~\ref{sec-imprintGW} we compute the imprint of a gravitational wave chirp in the stellar acoustic spectrum. In the last two sections we  present a discussion and conclusion about the impact and relevance 
of these results.

%%%%%%%%%%%%%%%%%%%%%%%%%%%%%%%%%%%%%%%%%%%%%%%%%%%%%%%%%%%%%%%%%%%%%%%%%%%%%
\section{The gravitation wave chirp Waveform}
\label{sec-GWS}
%%%%%%%%%%%%%%%%%%%%%%%%%%%%%%%%%%%%%%%%%%%%%%%%%%%%%%%%%%%%%%%%%%%%%%%%%%%%%

\begin{figure}
	\centering
	\includegraphics[scale=0.40]{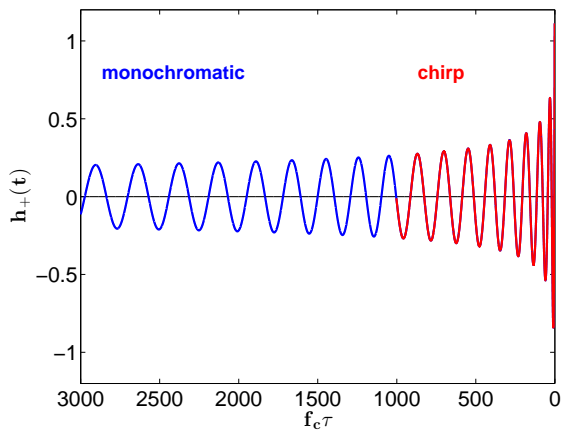}
	\caption{
		Strain of the gravitational wave of an inspiral stellar binary system:  $h_{+}(t)$ (with $h_\star=1$ and $g_k=1$) as a function of the dimensionless natural unity $f_c\tau$. This illustrates the two basic gravitational waveforms (or gravitational wave phases) of a inspiral binary system.  In the case that the time to coalescence is such that $f_c\tau \ge 1000 $ the wave is monochromatic. Reversely for  
		$f_c\tau \le 1000 $ the  gravitational wave is a chirp.}
	\label{fig:1}
\end{figure}

Binaries of compact stellar objects are the most studied sources of gravitational radiation. Among other binaries these apply to massive black-holes, EMRIs, stellar black holes, neutron stars, and white dwarfs. We start by assuming that the two compact stars of the binary system are in a circular orbit. 
The binary is considered to be sufficiently faraway from the {\it star detector}, such that the incoming gravitational radiation is  described by a plane wave field, but near enough to ignore the redshift corrections to the frequency spectrum. The gravitational waves  emitted in the end of the {\it inspiral phase} have a chirp waveform.  In this preliminary study,   we will compute gravitational radiation as due to the leading quadrupole which results from the use of  a multipole expansion, which also means that high-order terms will be of much smaller amplitude. This is valid since  $v/c\ll 1$. Hence, the binary system is held together by gravitational forces as per  the virial theorem  $(v/c)^2\sim (R_s/d)$ where $R_s$ is the Schwarzschild radius  and $d$ a typical length of the self-gravitating system, which 
leads to $R_s/d\ll 1$ for all the binaries systems in our study. For instance, for a binary like the one formed by the  supermassive black-hole and the S2 in the center of the  Milky Way, $( (R_s/d)\sim 0.0810A.U/12 A.U.\sim 0.0001 $. Nevertheless, more precise predictions of the waveform for gravitational radiation during the chirp phase can be found in the literature, such  in~\citep[][]{2002PhRvD..65f1501B}.  Unlike monochromatic waves, this waveform changes with time. The frequency and amplitude increase as follows:  

%%% 
\smallskip
(i) As the gravitational frequency $f$ is 2 times the orbital frequency, from Kepler's law results are that $f$ increases with the reduction of the orbital radius of the binary system.  The smaller star slowly enters in an adiabatic inspiral process by going through a succession of quasicircular orbits during which it loses energy by gravitational radiation. Consequently, $f$ increases as the time to coalescence $\tau$ decreases $f(\tau)/f_c=0.0728\left({f_c\tau}\right)^{-3/8}$  where $f_c$ is a characteristic frequency of the binary system. $f_c$ is equal to $c^3/(GM_c)$ where $c$ and $G$ are the speed of light and Newton constant,  and  $M_c$ is the chirp mass of the binary system  $M_c\equiv(m_1 m_2)^{3/5}/M_t^{1/5}$ where $m_1$ and $m_2$ are the masses of the two stars and $M_t=m_1+m_2$. 

%%%
The inspiral phase ends when the radial distance between the two stars is shorter than the last stable circular orbit, also known as the innermost stable circular orbit (ISCO). When this orbit is passed, the two stars merge and coalesce. The ISCO frequency $f_{ISCO}$ is approximately $2.2 kHz\; (M_\odot/M_t) $ where $M_\odot$ is the solar mass. Hence, the gravitational wave with the largest frequency emitted by the binary system  $f_{\rm max}$ at coalescence ($\tau=0$) is equal to twice the
$f_{ISCO}$~\citep{Maggiore:2008tka}. The precoalescence phase will be observed in the spectra of the {\it star detectors} if the $f_{max}$ of the binary has a value within the frequency interval of $10^{-7}$ to $10^{-2}$ Hz~\citep{2015ApJ...807..135L}, or $M_t$ has a value between $4.4\, 10^{5}$ and $4.4\, 10^{10}$  $M_\odot$.    
%%% 

\smallskip 
(ii) The strain $h$ increases as the binary system approaches the coalescence. The two polarized components of the $h$~\citep{Maggiore:2008tka}, $h_{+}$ and $h_{\times}$ are written in a condensed form, for the time interval corresponding to the orbital changes of the stars in the binary system, from the start of the inspiral phase until the coalescence ($- \infty <\tau <0 $): \begin{eqnarray} 
h_{k}(t)= h_\star \left(\frac{5}{f_c\tau}\right)^{1/4} g_{k}(\varphi)\; {\rm C}_{k} [\Phi\left(\tau\right)], \label{eq:htk} 
\end{eqnarray} 
where $k$ is one of the two possible polarizations $+$ or $\times$ and  $h_\star$ is the strain amplitude equal to $c/(d_\star f_c)$,  where $d_\star$ is the distance of the binary system to the {\it star detector}.  $g_{k}$ and  ${\rm C}_{k} $ are geometrical and  circular functions. The first is related with the direction of the gravitational wave source, and the latter takes into account the stretching of the gravitational wave as the binary approaches the coalescence. The $g_{k}$ functions are $g_{+}\equiv \left(1+\cos^2\varphi\right)/2 $ and $g_{\times}\equiv  \cos{\varphi}$ where $\varphi $ is a directional angle.  The ${\rm C}_{k} $ functions are $ C_{+}\equiv \cos{[\Phi\left(\tau\right)]}$  and $C_{\times}\equiv \sin{[\Phi\left(\tau\right)]}$.
These last functions are dependent of the phase  $\Phi\left(\tau\right)$, which is equal to $\Phi_o-2\left( f_c\tau/5 \right)^{5/8}$ where $\Phi_o$ is the value of the phase at coalescence. 
 
%%%
\medskip
The power spectrum of each of the $h_{k}(t) $ components, during the inspiral phase ($f \le f_{\rm max }$) is given by 
\begin{eqnarray} 
P_{k}(f)=\bar{h}_{\star}^2\;g_{k}^2(\varphi)\left(\frac{f}{f_c}\right)^{-14/6}, \label{eq:Pk} \end{eqnarray}  
where $\bar{h}_{\star}=A_s\; h_\star\tau_c$ with $\tau_c=f_c^{-1}$ and $A_s=0.2128$.
$\tau_c$ gives the time scale of the gravitational wave (GW) event, like for a burst or a Gaussian waveform;  
$P_{k}(f)$ is proportional to $h_\star^2\tau_c^2$~\citep{Maggiore:2008tka}.
The power spectrum $P_{k}(f)$ is equal to the square of the Fourier transform of  $h_{k}(t) $ (Eq.~\ref{eq:htk}), such that $ \tilde{h}_{k}(f)=\bar{h}_{\star}\;g_{k}(\varphi) \;e^{i\Psi_k(f)} \left({f}/{f_c}\right)^{-7/6} $ where the phase $\Psi_k(f)$ is either $ \Psi_+(f)=2\pi f(t_c+d_\star/c)-\Phi_o-{\pi}/{4}+{3}/{4}\left( 8\pi {f}/{f_c} \right)^{-5/3} $ or $\Psi_\times (f)=\Psi_+(f)+\pi/2$, e.g.,~\citep[][]{Maggiore:2008tka}. 

\begin{figure}
\centering
\includegraphics[scale=0.45]{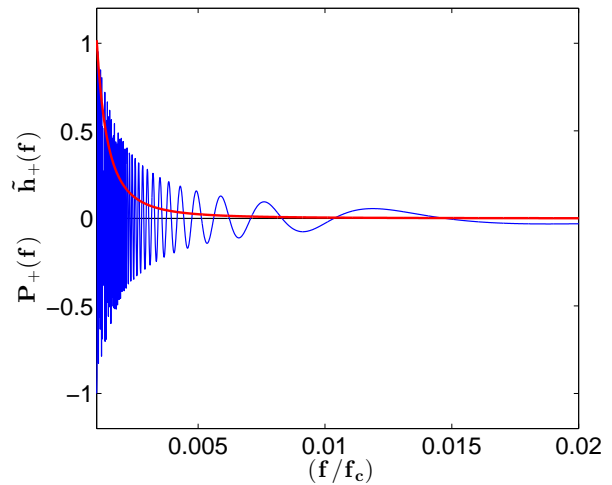}
\caption{Inspiral stellar binary system: $P_{+}(f)$ is 
the power spectrum of the strain in the chirp phase (red curve) and $\tilde{h}_{+}(f)$
the real part of the Fourier transform of $h_{+}(t)$ (blue curve),
as a function of the dimensionless natural quantity  $f/f_c$.   
Both functions are normalized to their maximum value.  
Figure~\ref{fig:1} shows the related $h_{+}(t)$.}
\label{fig:2}
\end{figure}

%%%
\medskip
Figure~\ref{fig:1} shows a gravitational wave with a positive polarization.
For large values of $f_c\tau$ the gravitational wave is monochromatic and for small values  of $f_c\tau$ the gravitational wave is 
a chirp. Equally, Fig.~\ref{fig:2} shows the power spectrum $P_{+}(f)$ and  $\tilde{h}_{+}(f)$ of the same gravitational wave. These results are easily generalized to $h(t)$ waveforms with both polarization, e.g.,~\citep[][]{2006PhRvD..73l2001T}. We notice that all gravitational waves produced by the different binary types have this waveform, the difference between them being uniquely related with the value of the characteristic frequency $f_c$ or their chirp mass. 

\begin{table} %[htb]
	\centering
	\caption{Inspiral Binary Systems}
	\resizebox{\columnwidth}{!}{%
		\begin{tabular}{llllllll}
			\hline
			\hline
			Binary  & $m_1\;\; m_2$  &  $M_c$      &  $f_c$ & $h_\star$\footnote{\scriptsize  The strain of a gravitational wave  at a distance of 1 kiloparsec from the binary.} & $f_{\rm max}$   \\
			$\qquad$ & ($M_\odot$)   & ($M_\odot$)  & ($Hz$) &  $\cdots$     &    ($Hz$)  \\
			\hline
			{\it low mass}    \\
			WD-WD& $0.71$ -- $ 0.13$  & $ 0.25$    & $8\,10^{5}$  & $1\,10^{-17}$ & $5\,10^{3}$\\
			%\hline
			NS-NS & $1.4$ -- $  1.4$        &  $1.22 $  & $2\,10^5$  &$6\,10^{-17}$& $2\;10^3$    \\
			%\hline
			BH-BH & $5.0$ -- $  5.0$        &  $4.35 $  & $5\,10^4$  &$2\,10^{-16}$& $4\,10^2$ \\
			%\hline 
			\\
			%\hline
			BH-BH & $14.2$ -- $  7.5$       &  $8.9 $  & $2\,10^4$ &$4\,10^{-16}$ & $200$   \\
			%\hline
			BH-BH & $36.0$ -- $  29.0$       &  $28.1 $  & $7\,10^3$  &$1\,10^{-15}$& $67$   \\
			%\hline
			%\hline
			{\it  high mass}    \\
			EMRI  & $4\;10^6$ -- $  15$          &  $2\; 10^3$ & $90$  &$1\,10^{-13}$& $1\;10^{-3}$   \\
			EMRI  & $\qquad\;$ -- $ 10^2$      &  $7\; 10^3$ & $29$  &$3\,10^{-13}$& $1\;10^{-3}$ \\
			EMRI  & $\qquad\;$ -- $ 10^3$      &  $3\; 10^4$ & $7$ & $1\,10^{-12}$& $1\;10^{-3}$  \\
			%\hline 
			\\
			%\hline
			BH-BH & $4\;10^6$  -- $ 10^4$      &  $1\; 10^5$ & $2$ &$5\,10^{-12}$& $1\;10^{-3}$   \\
			%\hline
			%\hline
			BH-BH & $10^5$ -- $  10^5$       &  $9\; 10^4$ & $2.3 $ &$4\,10^{-12}$& $2\,10^{-2}$   \\
			%\hline
			BH-BH &$5\,10^6$ -- $ \;5\,10^5$ &  $4\; 10^5$ & $0.46$  &$4\,10^{-11}$& $4\,10^{-3}$  \\
			%\hline
			BH-BH &$10^6$ -- $  10^6$        &  $9\; 10^5$ & $0.23$  &$2\,10^{-11}$& $2\,10^{-3}$     \\
			\hline
	\end{tabular}}
	\label{tab:1}
\end{table}

%%%%%%%%%%%
\medskip
Table~\ref{tab:1} lists the main characteristics of some typical inspiral binary systems.
In the table are included the two recently discovered stellar black-hole binaries~\citep{2016PhRvL.116f1102A,2016PhRvL.116x1103A}. 
Equally, we include EMRI binaries formed between the supermassive black hole of the Milky Way and 
several fiducial (stellar) black holes.    
Notice that as the chirp and total mass of the binary system increase, $f_c$ and $f_{\rm max}$ decrease. The chirp waveform is more pronounced for $f\le 0.005\,f_c$ (cf. Fig.~\ref{fig:2}).  In particular, for some binaries the duration of the chirp phase ($\tau f_c\le 1000$, cf. Fig.~\ref{fig:1}) varies between a few seconds to several minutes before coalescence, i.e., $\tau_\star$ is of the order of a few seconds to several minutes. This chirp phase occurs in a time scale
(or the equivalent frequency scale) that could be detected by sunlike stars, for which the spectral window of stellar oscillations varies from  $10^{-7}$ to $10^{-2}$ Hz. 
Therefore, binary systems that have chirp phases with a duration from a few seconds to several minutes,  
in principle, could affect the oscillations of nearby sunlike stars during this critical phase of their evolution.
Binaries of massive black holes should be the 
first candidates to consider, although these are unlikely to be found in the core of the Milky Way due 
to their very high masses (cf. Table~\ref{tab:1}). Nevertheless, EMRIs binaries that equally emit gravitational 
radiation in the same spectral window are much more likely to be found in the core of the Milky Way. 

%%%%%%%%%%%
\medskip
During the chirp phase several quadrupole  modes in the same star will be excited by the passing gravitational wave. Conversely, in the case of monochromatic gravitational waves, at best 
a single stellar quadrupole mode will be excited (cf. Fig.~\ref{fig:1}). It is worth highlighting that as the inspiral binary approaches the coalescence, $h(t)$ is less accurate and the determination of the exact waveform should take into account the finite structure of the two compact objects in coalescence. Nevertheless the general properties of the gravitational waveform remain.

%
%%%%%%%%%%%%%%%%%%%%%%%%%%%%%%%%%%%%%%%%%%%%%%%%%%%%%%%%%%%%%%%%%%%%%%%%%%%%%
\section{Imprint of gravitational waves on the acoustic spectrum}
\label{sec-imprintGW}
%%%%%%%%%%%%%%%%%%%%%%%%%%%%%%%%%%%%%%%%%%%%%%%%%%%%%%%%%%%%%%%%%%%%%%%%%%%%%

%%%%%%%%%%%
Main-sequence stars are in hydrostatic equilibrium, their transport of energy and their mechanical vibrations occur in quite distinct time scales, the Kelvin-Helmholtz and free-fall time scales. In the Sun these characteristic times are of the order of 30 million years and 40 minutes respectively. For that reason, the stellar vibrations are represented as a combination of nonradial adiabatic oscillation modes of low amplitude, e.g.,~\citep[][]{1989nos..book.....U}. 
Any perturbed quantity of a mode, like the displacement $\xi_n ({\bf r},t)$
is equal to $A (t) \xi_n ({\bf r})\;e^{-i\omega_{\rm n} t-\eta_{\rm n}t}$, where $\omega_{\rm n} $  and $\eta_{\rm n} $ are the frequency and the damping rate,
$A(t)$ is the instantaneous amplitude, and $\xi_n ({\bf r})$ is the spatial eigenfunction, e.g.,~\citep[][]{2005MNRAS.360..859C,2001A&A...370..136S,2001A&A...373..916L,1980tsp..book.....C}.  The instantaneous amplitude is a solution of 
\begin{eqnarray}  \frac{d^2A}{dt^2} +2\eta_{n} \frac{dA}{dt} +\omega_n^2 A =  {\cal S_{\rm gw } } (t),  \label{eq:Amotion}  \end{eqnarray} where ${\cal S_{\rm gw } }(t)$ is the source of gravitational radiation~\citep[e.g.,][]{1973grav.book.....M}. Following from the specific properties of gravitational systems as  demonstrated in general relativity, a gravitational perturbation only affects modes with degrees equal to or larger than 2. For convenience, we opt to study the leading order of the gravitational perturbation on the quadrupole acoustic modes, see~\citep[][]{LopesSilkGW2014} for details. In this study the mode $p_n$ ($n=0,1,\cdots$) 
refers to  $f$- and acoustic ($p$-) quadrupole modes of order $n$ and undefined azimuthal order $m$, e.g.,~\citep[][]{Maggiore:2008tka}.  
${\cal S_{\rm gw } }(t)$ is equal to the  $ L_{n}\; \ddot{h}_{m}(t) $ where $L_{n}$ is the modal length of the quadrupole mode and $\ddot{h}_{m}(t)$ is the $m$-spherical component ($m\le 2 $) of perturbation of spatial component of the minkowski metric, e.g.,~\citep[][]{1973grav.book.....M}.
In this study, we assume that the source of gravitational radiation
is at a sufficiently small distance from the {\it star detector}, such that we can neglect all redshift corrections. This is justified since all the gravitational sources are located in our own Galaxy.
 $L_{n}$ is equal to $1/2\; R\;|\chi_{n}|$ where $R$ is the stellar radius and $\chi_{n}$ reads 
\begin{eqnarray} \chi_{n}=\frac{3}{4\pi \bar{\rho}_{\star}} \int_0^1 \rho (r) \left[ \xi_{r,n2}(r)+3  \xi_{h,n2}(r) \right] r^3 dr, \label{eq:chin} \end{eqnarray} 
where $\rho$ and $\bar{\rho}_{\star} $ are the density inside the star and its  averaged value.
This definition of  $\chi_n$ is identical to the one used for resonant-mass detectors~\citep{LopesSilkGW2014}. In the particular case that  $\bar{\rho}_{\star}$ is constant,  Eq.~(\ref{eq:Amotion}) becomes equivalent to the one found for a  spherical resonant-mass detector,  e.g.,~\citep[][]{Maggiore:2008tka}.  We noticed that there are several $\chi_n$ definitions, e.g.,~\citep[][]{1984ApJ...286..387B,2011ApJ...729..137S} differing between them only by the normalization condition.

%%%%%%%%%%%%%%%%%%%%%%%%%%%%%%%%%%%%%%%%%%%%%%%%%%%%%%%%%%%%%%%%%%%%%%%%%%%%%
% // TABLE  //
%\input{table_nusun}
%\input{table_nuQ_star}
\begin{table} %[htb]
	\centering
	\caption{Quadrupole Acoustic Modes  of One Solar Mass}
	\begin{tabular}{lllll} 
		\hline
		\hline
		Mode
		&  Frequency~\footnote{Frequency table of solar acoustic modes  obtained from  a compilation made by~\citet{2012RAA....12.1107T}. The frequencies in italic correspond to theoretical predictions for the current solar model as in reference~\citet{2013ApJ...765...14L}.} &      $L_n$ &  $Q_n$    \\
		&  ($Hz$) &   $({\rm cm })$ &  ({\rm no-dim}) \\
		% Table 1 - Turck-Chieze & Lopes ilidio
		\hline
		&   $\times 10^{-6}$  &       $\times 10^{7}$&  $\times 10^{+8}$  \\
		$f$   & $\mathit{347}$   & $2.347$  & $38$ \\
		$p_1$ & $\mathit{382}$  & $3.841$  &  $4.1$  \\
		$p_2$ & $\mathit{514}$ & $0.737$  &  $1.0$  \\
		$p_3$ &  $\mathit{664}$ & $0.219$ &   $0.28$  \\
		$p_4$ & $\mathit{811}$ & $0.074$ & $0.10$ \\
		\hline 
		&   $\times 10^{-6}$       &      $\times 10^{5}$   &  $\times 10^{6}$  \\
		$p_5$ & $\mathit{959}$  &$2.867$ &  $3.8$   \\
		$p_6$ & $\mathit{1104}$ &$1.211$ &  $1.7$  \\
		$p_7$ &$\mathit{1249}$   & $0.524$ &  $0.74$  \\
		$p_8$& $1394$  &  $0.238$& $0.33$ &    \\
		$p_9$& $1535$  & $0.108$ &  $0.12$ &   \\
		\hline 
		&     $\times 10^{-6}$      &     $\times 10^4$ &  $\times 10^{4}$    \\ 
		$p_{10}$ & $1674$    &    $1.082$ &   $6.7$ \\
		$p_{11}$ & $1810$    &     $0.520$ &    $4.1$\\
		$p_{12}$ & $1945$   &     $0.272$ &  $2.9$  \\
		$p_{13}$ & $2082$     &    $0.153$&    $2.1$  \\
		$p_{14}$ & $2217$      &    $0.054$&    $1.9$  \\
		$p_{15}$ & $2352$      &     $0.033$&  $1.8$  \\
		$p_{16}$ & $2485$    &      $0.022$&  $1.7$  \\
		$p_{17}$ & $2619$       &      $0.014$&  $1.5$  \\
		$p_{18}$ & $2754.$     &    $0.010$&  $1.5$ \\ 
		\hline
		\hline
	\end{tabular}
	\label{tab:nuQstar}
\end{table}   
%%%%%%%%%%%%%%%%%%%%%%%%%%%%%%%%%%%%%%%%%%%%%%%%%%%%%%%%%%%%%%%%%%%%%%%%%%%%%

%%%%%%%%%%%
\smallskip
In the computation  of the stellar acoustic modes and $\chi_{n}$,  we use an up-to-date solar model, e.g.,~\citep[][]{2013MNRAS.435.2109L}. 
The observational frequencies and  damping rates used were from
~\citet{2000ApJ...537L.143B,2001SoPh..200..361G,2004ApJ...604..455T,2009ApJS..184..288J} 
and~\citet{1997MNRAS.288..623C, 2005A&A...433..349B}. 
The theoretical damping rates used were from~\citet{1999A&A...351..582H,2005A&A...434.1055G,2013ASPC..479...61B}. 
Table~\ref{tab:nuQstar} shows the frequencies for the quadrupole acoustic modes of a main sequence star that we chose to be identical to the Sun. The frequency in Hz in the table corresponds to $\omega_n/2\pi$. A detailed discussion about the properties of low-order acoustic modes can be found in~\citet{2012RAA....12.1107T}. 

%%%%%%%%%%%

\smallskip
The averaged power spectrum $P_{A}$ ($=\langle |\tilde{A}^2| \rangle$) of an acoustic mode  stimulated by a gravitational wave [see Eq.~\ref{eq:Pk})] is computed by taking the Fourier transform of Eq.~(\ref{eq:Amotion}) and neglecting transients terms arising from the initial conditions on $A$.  $P_{A}$ of an quadrupole mode reads
\begin{eqnarray}  
P_{\rm A}(\omega) =  \frac{L_{n}^2\;\bar{h}_{\star}^2\;\omega_c^4\;g_{k}^2(\varphi)}{(\omega^2-\omega_{n}^2)^2+4\eta_{n}^2\omega^2} \left(\frac{\omega}{\omega_c}\right)^{5/3}, \label{eq:Pfgw} 
\end{eqnarray}
where $\omega_c=2\pi f_c$. 

%%%%%%%%%%%
\smallskip
The square of photospheric velocity $V^2 (\omega)$ is equal to  $(2\pi \tau_{\rm w})^{-1} \int_{0}^{+\infty} P_{\rm   V}(\omega)d\omega$, where $ \tau_{\rm w}$ is the duration of the gravitational wave impact on the star's quadrupole mode. 
$P_{\rm V}$ ($\equiv \omega^2P_{\rm A}$) is the power spectrum of the square of photospheric velocity~\citep{2005A&A...433..349B}.
%%%%%%%%%%%
Hence  the impact of a {\it chirp gravitational wave emission}  [as defined by Eqs.~\ref{eq:htk} and~\ref{eq:Pk}] with a frequency $\omega $ that resonates with the  frequency of the stellar mode $\omega_n$  reads 
\begin{eqnarray}
V_n^2 =
\frac{h_\star^2 L_{n}^2\omega_n^4 }{\alpha_s^2\eta_n^2}\;
{\cal C}_n,
\label{eq:Vnf}
\end{eqnarray}
where $V_n^2$ is  equal to $V^2(\omega_n)$, ${\cal C}_n$ is a chirp factor given by  $ \left(\tau_c^2 \Delta \omega_n/\tau_{\rm w} \right) \left(\omega_n/\omega_c\right)^{-7/3}$, and $\alpha_{\rm s}$ ($\equiv (2\sqrt{2\pi})/(A_s g_{k}(\varphi)\gamma_s)$) 
is a multiplicative factor related to stellar observations.
$\alpha_{\rm s}$ (with $g_{k}=1$) is equal to $24/\gamma_s $ where $\gamma_s$ is a unity photospheric numerical factor.
As the contribution for the $V^2(\omega)$ integral is only significant near each  $\omega_n$, in computing the previous equation, we approximate $\omega^{11/3}((\omega^2-\omega_{n}^2)^2+4\eta_{n}^2\omega^2)^{-1}$ in $P_{\rm V}(\omega)$ by its first term of the Taylor series $\omega_n^{5/3}/4\eta_n^2$, and the integral limits around each $\omega_n$ by
$\omega_n-\Delta\omega_n/2$ and $\omega_n+\Delta\omega_n/2$
where  $\Delta\omega_n$  is the equivalent linewidth. We notice that $V^2(\omega_n)$ is proportional to $\eta_n^{-2}$:   
Eqs.~(\ref{eq:Pfgw}) and~(\ref{eq:Vnf}) are related  with $V^2 (\omega_n)=\omega_n^2(\Delta\omega_n/\tau_{\rm w})/2\pi\;P_{A}(\omega_n)$. Thus, if the linewidth $\Delta\omega_{n}$
relates with the damping time as  $\Delta\omega_{n}\sim \tau_{\eta}^{-1}$,
and  $\tau_{\rm w}\sim \tau_c $ then $ \left(\tau_c^2 \Delta \omega_n/\tau_{\rm w} \right)$ in ${\cal C}_n$  simplifies to $\tau_c/\tau_{\eta}$.

\begin{figure}
	\centering
	\includegraphics[scale=0.45]{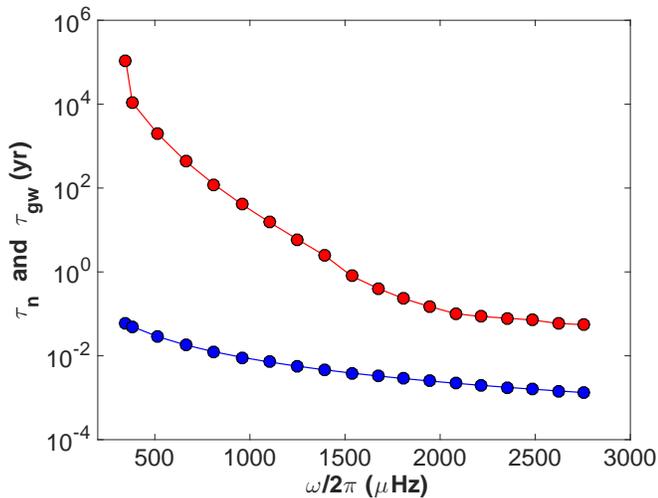}
	\caption{Comparison between the dumping time of the quadrupole acoustic modes 		$\tau_n$({\it red circle})  and the characteristic time drift time of the gravitational  waves $\tau_{gw}$({\it blue circle}).The  gravitational radiation is assumed to be due to  the occurrence of an  EMRI (with $m_1=4\;10^6\; M_\odot$ and $m_2=15\;\; M_\odot$)  at the center of the Galaxy  (see Table~\ref{tab:1}).}
	\label{fig:3}
\end{figure}

\smallskip
In the particular case that the excitation of the stellar mode is due to	  
{\it monochromatic gravitational wave emission}, it is reasonable to assume 
that $\omega_n\sim \omega_c$ and  $\tau_{n}\sim \tau_{c}$ for which ${\cal C}_n=1$.  In these circumstances Eq.~(\ref{eq:Vnf}) simplifies to
$ V_n={h_{\star}L_{n}\omega_n^{2}}/{(\alpha_{s}\eta_n)} $.
Equally  $ V_n$ can be expressed as a function of the quality factor of the mode 
$Q_n\equiv \omega_n/(2\eta_n)$ as $ V_n=2{h_{\star}L_{n} Q_{n} \omega_n}/{\alpha_{s}} $.	
The values of $Q_n$ are shown in Table~\ref{tab:nuQstar}.
This result was  computed by~\citet{LopesSilkGW2014}, and an identical expression was previously obtained by~\citet{2011ApJ...729..137S} for  acoustic
and gravity quadrupole modes in the Sun. Moreover, this result shows that the photospheric  velocity of modes  excited by a gravitational chirp waveform [Eq.~\ref{eq:Vnf} with ${\cal C}_n\neq 1$] differs only by the simple factor from the excitation due to a monochromatic gravitational wave.  

\smallskip 
In both gravitational emission cases  the excitation of a quadrupole mode occurs when the frequency of the gravitational wave $\omega$ 
matches the frequency of the stellar mode $\omega_n$. Nevertheless,
this only occurs for a very short period of time; hence, the exact calculation of the impact of the gravitational waves
must take into account the time drift of $\omega$ in relation to $\omega_n$. In the case that the time drift is small, 
$\omega$ is approximated by $\omega_n+\dot{\omega}\tau$, where  $\dot{\omega}$ is the frequency time variation of the gravitational
wave  and  $\tau$  is the time difference to the resonance~\citep{2015ApJ...807..135L}.

\smallskip

The impact of the gravitational wave on an oscillation mode is maximum when the gravitational time drift $\tau_{gw}\equiv 1/\sqrt{\dot{\omega}}$  is  larger than the damping time 
of the mode $\tau_n=1/(2\eta_n)$, i.e., the ratio ${\cal T}_n=\tau_{gw}/\tau_n$ is larger than one. 
Figure~\ref{fig:3} compares these two characteristic times.
In this study   $\tau_{gw}$  is 
$0.577\left(GM_c/c^2 \right)^{ -5/6} \omega^{-11/6} $, e.g.,
~\citep[][]{2015ApJ...807..135L,Maggiore:2008tka}.   
Accordingly, a stellar mode for which  ${\cal T}_n\gg 1$, the excitation
is known  as the steady-state solution or a {\it saturated mode of oscillation}.
In the calculation of this photospheric velocity $V_{n,s}$, 
the frequency drift $\dot{\omega}$  is neglected,  as such  $V_{n,s}=V_n $.
Reversely, for the case that ${\cal T}_n\ll 1$,  the excitation is known as an {\it undamped mode of oscillation}, the  photospheric velocity $V_{n,u}$  is such that  the contribution $\dot{\omega}$ is taken into account in the calculation of   $V_{n,u}$, following ~\citet{2015ApJ...807..135L} and~\citet{2014MNRAS.445L..74M}  $V_{n,u}={\cal T}_n^{1/2} V_n $.
Figure~\ref{fig:4} shows the $V_{n,u}$  of 
quadrupole acoustic modes excited by  a {\it chirp gravitational wave emission} and a {\it monochromatic gravitational wave emission}. In this case  all stellar modes  have a ${\cal T}_n $  that varies from $10^{-6}$ to $10^{-2}$; therefore, 
all the modes are in an undamped mode of oscillation since ${\cal T}_n\ll 1$.
$V_{n,s}$ is also shown in the same figure. In the calculation of $V_{n,s}$  and $V_{n,u}$  we used the Eq.~(\ref{eq:Vnf}) for which the gravitational wave radiation was estimated from the Eq.~(\ref{eq:Pk}). The power spectrum of the incoming gravitational radiation is given by Eq.~(\ref{eq:htk}).

%%%%%%%%%%%%%%%%%%%%%%%%%%%%%%%%%%%%%%%%%%%%%%%%%%%%%%%%%%%%%%%%%%%%%%%%%%%%%
\section{Discussion}
%%%%%%%%%%%%%%%%%%%%%%%%%%%%%%%%%%%%%%%%%%%%%%%%%%%%%%%%%%%%%%%%%%%%%%%%%%%%%

%%%%
The most important factor affecting the amplitude of $V_n$	
is the distance of the {\it star detector}	to the compact binary. 
An EMRI merger corresponding to the capture of the S2 star by the supermassive black hole in the Galactic Center will produce a gravitational event with a strain amplitude of $\sim 10^{-13}$ at  a distance of 1000 parsecs. This result is obtained  from the relation $h_\star=c/(d_\star f_c)$ (see previous section) where $f_c=90 Hz$ (see Table~\ref{tab:1}). In particular, for the case of  a  binary of black holes that has masses identical to the ones found for the first time by the LIGO Collaboration~\citep{2016PhRvL.116f1102A}, the strain amplitude at the same distance is $\sim 10^{-15}$ (see Table~\ref{tab:1}).
This corresponds to  a strain of  $ 3\; 10^{-21}$ at a distance of 410 Mpc,
which is close to  the $10^{-21}$ strain amplitude  measured by the LIGO experiment on Earth.	

\begin{figure}
	\centering
	\includegraphics[scale=0.45]{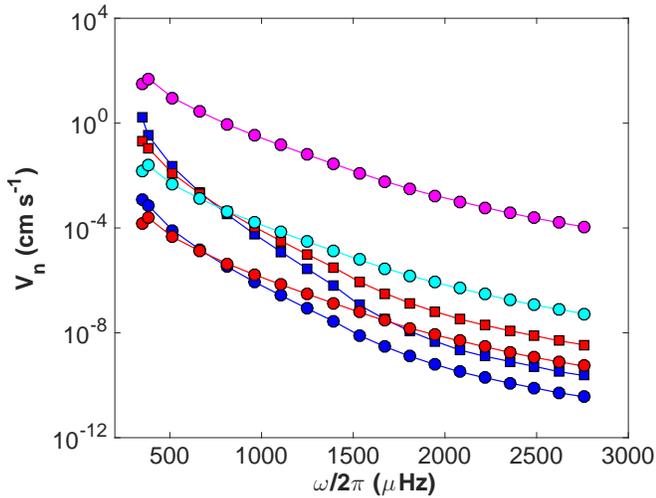}
	\caption{
		Photospheric velocity amplitude of the low-order acoustic quadrupole modes ($n=0,1,\cdots$) of a {\it star detector} located at a distance of 1 kiloparsec from the Galactic Center for two types of gravitational radiation emission:  {\bf (a)}
		{\it monochromatic wave}, $V_{n,s}$ ({\it blue square})  and  $V_{n,u}$ ({\it blue circle})
		as given by Eq.~(\ref{eq:Vnf}) with ${\cal C}_n=1$. 
		{\bf (b)} {\it  chirp wave },
		$V_{n,s}$ ({\it red square})  and  $V_{n,u}$ ({\it red circle})
		as given by Eq.~(\ref{eq:Vnf}) with ${\cal C}_n\neq 1$. 
		{\bf (c)}  $V_{n,u}$  is the same used in case (b), but now
		the {\it star detector} is located at a 10 parsec
		({\it cyan circle}) or 1000 A.U. ({\it magenta circle}).
		The  gravitational radiation is assumed to be due to  the occurrence of an 
		EMRI (with $m_1=4\;10^6\; M_\odot$ and $m_2=15\;\; M_\odot$)  at the center of the Galaxy  (see Table~\ref{tab:1}).}
	\label{fig:4}
\end{figure}

Figure~\ref{fig:4} shows the $V_{n,u}$ for this EMRI event, where the 
sun-like star is located at a distance of 1000 parsec,  10 parsecs  and 1000 A.U. from the Galactic cCnter. The last distance, although unlikely, gives us an order of magnitude of the phenomena.  For illustrative purposes other quantities are also shown in the same figure.
The Sun is the fiducial {\it star detector} in this analysis. In the computation of the photospheric velocity, the values of $\omega_n$ and $\eta_n$ correspond to the observed solar frequencies and the  theoretical predictions of damping rates for the Sun by~\citet{2013ASPC..479...61B} and~\citet{1999A&A...351..582H}.
The $L_n$ varies from $10^7$ -- $10^4$ cm as computed by~\citet{2015ApJ...807..135L} for these modes.  $\alpha$ has the numerical value $24.0$ (with $\gamma_s\sim 1$). 

These results can be understood qualitatively. If we neglect
numerical factors of the order of unity, the estimation of $V_{n,u}$ is made as follows: 
for a monochromatic gravitation wave emission [from the expression given by Eq.~(\ref{eq:Vnf}) with ${\cal C}_n=1$], we obtain that $V_{n,u}$  is proportional to $h_\star L_n Q_n {\cal T}^{1/2}_n$ where $Q_n$ is the quality factor of the quadrupole mode of order $n$, such that  $Q_n=\omega_n/(2\eta_n)$.  If we choose an $\eta_n \sim 10^{-6}\;\mu Hz$ as a fiducial value of the range of $\eta_n$ values $10^{-8}$ -- $10^{-3}\;\mu Hz$  predicted by \citet{2013ASPC..479...61B} and~\citet{1999A&A...351..582H} for low-order acoustic modes, 
for a mode with a frequency $400 \mu Hz$ we obtain $Q_n\sim 10^9$. 
Note that a $Q_n$ computed from an observational data set
(acoustic modes with much higher frequency), we obtain  significantly smaller  $Q_n$  values. A typical example from \citet{1997MNRAS.288..623C}, corresponds to a mode with $\nu\sim 1500 \mu Hz$ and $\eta_n \sim 10^{-2} \mu Hz$  for which $Q_n\sim 10^{6}$ .  Nevertheless, this result can only be used as a lower value estimation of $Q_n$, since acoustic modes with these high frequencies (and much higher values  of $\eta_n$) are not perturbed by incoming gravitational radiation. Therefore, for a fiducial acoustic mode with a frequency  of $400 \mu Hz$, we estimate   $V_n\sim$ 1  cm s$^{-1}$ for  a $Q_n\sim 10^{9}$, $L_n\sim 10^7$ cm and ${\cal C}_n=1 $   when stimulated  by an incoming gravitational wave with a strain  $h_\star=10^{-13}$. Finally, if we take into account that
this mode is unsaturated, this value must be multiplied by the ${\cal T}^{1/2}_n$  with $\tau_{gw}\sim 10^{6}$ s and $\tau_n\sim 10^{12}$ s  than $V_{n,u}\sim 10^{-3}$ cm s$^{-1}$ (Cf. Figure~\ref{fig:4}). However, if the {\it star detector} is located at a distance of $\sim 1000$ AU then  $V_n\sim 10$ cm s$^{-1}$. This study complements the original work of~\citet{2014MNRAS.445L..74M}, which has found  stars to be good resonant absorbers of gravitational radiation. The contribution related with the chirp emission is contained in the term ${\cal C}_n$ [Eq. (\ref{eq:Vnf}] with ${\cal C}_n\neq 1$). This quantity increases with the frequency varying from $10^{-2}$ up to  $10^{2}$. Accordingly,  $V_{n,u}$ of low-order modes excited 
by a chirp gravitational wave (${\cal C}_n\neq 1 $) is a factor 10
smaller in comparison to modes excited by a monochromatic wave
(${\cal C}_n=1 $), since their velocity ratio is proportional to  ${\cal C}^{1/2}_n$ (cf. Fig.~\ref{fig:4}). 
  
In the following, we discuss the two  $V_{n,u}$ results given equation~(\ref{eq:Vnf}) with ${\cal C}_n=1$  and ${\cal C}_n\neq1$.  These solutions correspond to a
monochromatic emission and a  chirp emission  of gravitational radiation. In both cases, the amplitude of $V(\omega_n)$ decreases with the order mode $n$, since  the modal length $L_n$ (and $\chi_{n}$)  decreases rapidly with increasing $n$ acoustic modes for a main-sequence star.  In the following 
it is worth two highlighting the following:

%%%%
\smallskip
First,  the $V_{n,u}$ for the low-$n$ quadrupole acoustic modes is of the order of  $10^{-4}$ -- $1\;{\rm cm\; s^{-1}}$ (cf. Fig.~\ref{fig:4}, depending of on the distance of the {\it star detector} to the binary). These values are below the $V_{n,u}$ currently measured for similar stars in the neighborhood of the Sun by the {\sc Kepler} mission, for which the excitation of stellar oscillations is well known to be attributed to the convection of the external layers of these stars. As an example, the Procyon A star (F5 IV spectral type star) 
has $V_n$  $\sim38\;{\rm cm s^{-1}}$, e.g.,~\citep[][]{2010ApJ...713..935B}.
This result is equally valid for monochromatic and chirp emission phases of the inspiraling binary.  

\begin{figure}
	\centering
	\includegraphics[scale=0.45]{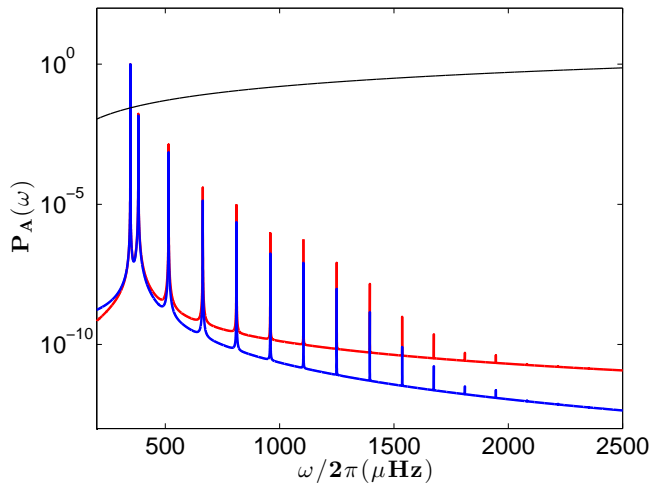}
	\caption{Amplitude power spectrum of the quadrupole modes of different orders excited by an external GW source
		with a characteristic frequency $f_c\approx 90 Hz$ (identical to a inspiral binary system with $m_1=4\;10^6\; M_\odot$ and $m_2=15\;\; M_\odot$, see Table~\ref{tab:1}):  the peaks occur at the location of eigenfrequencies $\omega_n$ ($l=2$ and $n=0,1,2,3,\cdots $) corresponding to the different acoustic eigenmodes of the Sun. The red curve corresponds to the amplitude power spectrum (Eq.~(\ref{eq:Pfgw})). 
		The blue curve corresponds to the amplitude power spectrum as given by Eq.~(\ref{eq:Pfgw}) with the term $(\omega/\omega_c)^{5/3}$ replaced by one. The black curve corresponds to the term $(\omega/\omega_c)^{5/3}$. 
		All the curves are scaled by their maximum values (in arbitrary units). }
	\label{fig:5}
\end{figure}

%%%%
\smallskip
- Second, the impact of the gravitational waves during the chirp emission phase on  $ V_{n,u}$ [Eq.~\ref{eq:Vnf}] is strongly dependent on the shape of the strain function $h(t)$ [Eq.~\ref{eq:htk}]. Unlike for the case of excitation of $V_{n,u}$ by a monochromatic gravitational wave 
for which only a stellar quadrupole mode is excited, during the chirp phase  
several acoustic modes are excited sequentially  by the same gravitational waveform.   Figure~\ref{fig:5} shows how the global shape of the $P_A(\omega)$  spectrum for quadrupole acoustic modes in the sunlike star is excited by the gravitational radiation coming from the inspiral binary during the chirp emission phase. The $P_A(\omega)$  corresponds to a gravitational event shown in Fig.~\ref{fig:5} and the $ V_{n,u}$ is given by Eq.~(\ref{eq:Vnf}).

 \smallskip
These values of $V_{n,u}$ predicted for  stars similar to the Sun near the galactic core  should be within reach only in a future  generation of asteroseismology satellites. Moreover, as  $V_{n,u}$ decreases with $d_\star$, it is reasonable to expect that  {\it star detectors} located in the neighborhood of such binaries (as near as 1000 AU), could have a $V_n$ above the threshold of detectability. 
However for larger $d_\star$ values, such phenomena
will be difficult to observe.  In particular, it is unlikely  for that  measurement to be 
done by  the {\sc PLATO} mission~\citep{2013EGUGA..15.2581R},
since, at best  {\sc PLATO}   is expected to measure oscillations
in sunlike stars with amplitudes of the order of $\sim 1\;{\rm cm\;s^{-1}}$.
It will be necessary to wait for an increase of at least one or two orders of magnitude  in the instrumental threshold  to be able to measure  quadrupole acoustic modes excited  by  gravitational radiation coming from the Galactic Center. This point is illustrated
 in Fig.~\ref{fig:3} where it is shown the  $V_{n,u}$ of a gravitational chirp emission
 of a star detector located at distances of 10 parsec and 1000 A.U. for which
 the   $V_{n,u}$ increases to $10^{-2}$ cm s$^{-1}$ and $50$  cm s$^{-1}$, respectively. 
Although the last scenario is theoretically possible, it will be very unlikely since the Schwarzschild radius of the supermassive black hole is  0.0810 A.U. and the orbit of the S2 varies between 12 and 2000 AU. It would mean that such star would also be orbiting the supermassive black hole.
For comparison, it is worth noticing that in the Sun's case, the precision attained in $V_n$ by the {\sc GOLF} experiment for a 10 year observational period~\citep{2004ApJ...604..455T,2007Sci...316.1591G,2009ApJS..184..288J} varies from $10^{-2}{\rm cm\;s^{-1}}$ to $3\,10^{-4}{\rm cm\;s^{-1}}$. The signal-to-noise ratio of the {\sc GOLF} experiment is just a few orders of magnitude below the $V_n$ predictions previously mentioned.  
 %CLAUDIA

%%%%%%%%%%%%%%%%%%%%%%%%%%%%%%%%%%%%%%%%%%%%%%%%%%%%%%%%%%%%%%%%%%%%%%%%%%%%%
\section{Conclusion}
%%%%%%%%%%%%%%%%%%%%%%%%%%%%%%%%%%%%%%%%%%%%%%%%%%%%%%%%%%%%%%%%%%%%%%%%%%%%%

%%%%
Main-sequence stars like the Sun (with a spectral window of acoustic oscillations  $300 \mu Hz \le \nu_n \le 5000 \mu Hz$) when located at relatively short distances of compact binaries (including massive black-hole and EMRIs binaries) of the Milky Way core have their quadrupole acoustic modes of low order stimulated by the incoming gravitational  radiation. 
This frequency  range overlaps the frequency window of
gravitational waves emitted by EMRIs  $100 \mu Hz \le \nu_n \le 10000 \mu Hz$. These systems form preferentially in dense stellar regions such as the nucleus of galaxies. The Galaxy nucleus  is on of the most dense stellar regions in the Universe  with $\sim 10^7$ stars squeezed in spherical regions with radius of $\sim 10$ parsec~\citep{2014ApJ...784...23M}.  As in any other galaxy,
the nucleus of the Milky Way is one of the preferential locations to look for EMRIs. 
In particular, these {\it detector's stars} could follow the end of 
the binary contraction in the pre-coalescence phase, during which 
sequentially the low-$n$ quadrupole acoustic modes of the star 
are stimulated by the incoming  gravitational waves.   
Equally, many other stars, including  main-sequence, subgiant and red giant stars~\citep{2013sf2a.conf...25M} 
will also be sensitive to the same type of radiation. 
Hence, all these sunlike stars have a combined spectral window of  $0.1\;  \mu Hz$  to $10^{5}\;  \mu Hz$. As such these stars form a network of detectors sensitive to the gravitational radiation coming from the Galactic Center.

\smallskip
A very interesting result of this study is the  clear possibility to observe  the end phase of the 
coalescence of binary systems by using sunlike stars as detectors.    
This is a powerful method to study the gravitational radiation.
As the period of the gravitational wave chirp varies 
within the spectral bandwidth of the {\it star detector}, it is certain that 
different quadrupole modes of the same star will register the same gravitational event. 
Moreover, as $V_n (\omega)$ is proportional to $(\omega/\omega_c)^{-7/6}$, this relation
can be used to look for the gravitational wave signature on low-order  quadrupole modes of these stars
(cf. Eq.~\ref{eq:Vnf}).
This method is ideal for studying the  binaries of massive black holes or EMRIs 
for which the  chirp phase occurs  in a time interval varying from a few seconds to several minutes (cf. Table~\ref{tab:1}).   
This new type of research can complement the gravitational waves experimental detectors   like the ELISA instrument. 
Moreover these {\it star detectors} can be used to look for gravitational wave radiation, including  the chirp phase of inspiral binary systems in the frequency interval, $10^{-6}$ to $10^{-4}$ Hz, which is not currently probed by ground based experiments.                 

%%%%%%%%%%%
% IPL
%%%%%%%%%%%
\smallskip
\smallskip

The  galactic core is a very efficient machine for converting the gravitational energy of the 
captured matter into electromagnetic and (possibly) gravitational radiation. 
Stars like the Sun but near the galactic center, for which their spectra of oscillations is well known, 
form a natural network of detectors for gravitational radiation. In this article, we have shown for the first time that chirp waveforms 
of gravitational waves have a unique imprint in the spectrum of these sunlike stars. 
Nevertheless,  it is worth highlighting that this study is made for a relatively simple chirp
waveform expression although  sufficient to make the first prediction of the amplitude
stellar modes. A more rigorous calculation must take into account an high-order expression gravitational-wave emission during the inspiral of compact binary systems beyond
the  quadrupole radiation expression~\citep{2002PhRvD..65f1501B}.

\smallskip
We could expect that it would be quite difficult to separate the excitation of quadrupole modes in sunlike stars caused by gravitational radiation from the intrinsic excitation and damping of these modes due to  turbulent convective motions occurring on the upper-layers of the star. 
Nevertheless, there are three important arguments that could help astronomers to isolate the gravitational wave stimulation from intrinsic excitation:
First, current theory of stochastic excitation and damping of nonradial oscillations is very successful in predicting the exactly amplitude of individual acoustic modes, e.g.,~\citep[][]{2007A&A...463..297S}, as well as the global envelope of amplitudes of the acoustic modes in the oscillation spectrum of the star, e.g.,~\citep[][]{2008ApJ...682.1370K}. 
Second, this theory predicts that low-order degree modes (including radial, dipole and quadrupole modes) with near frequencies have identical amplitudes. Therefore, by taking advantage from the fact that gravitational waves only excite quadrupole modes, the amplitude excess found in these modes above the radial and dipole mode amplitudes can be attributed to excitation due to gravitational radiation. 
This means that if astronomers found on the oscillation spectrum of a sunlike star one or more quadrupole modes with amplitudes that are well above the amplitudes of neighbouring radial and dipole modes, this will be a strong indication that these modes are being stimulated by incoming gravitational radiation, possibly caused by a source located nearby the star.  
Finally, if the gravitational source is known, it will be possible to compute precisely the amplitude of each mode due to the impact of the gravitational wave,
in particular by taking into account the distance and direction of the gravitational source in relation to the star~\citep{LopesSilkGW2014}.  Although,  gravitational waves affect all modes with a degree higher than two, the amplitudes of high-degree modes are very small in comparison to quadrupole modes. Actually, this effect is neglected in high-degree modes  since  gravitational wave stimulation is insignificant. 

\smallskip
As the stars located near the core of the Milky Way have their line of sight obscured by dust, the near infrared band will provide the best option to observe such stars. In principle, a near infrared observatory should be able to observe stars in these dense stellar regions of the galactic nucleus.  This option could be a very interesting alternative to optical asteroseismolgy, since the amplitude of stellar oscillations in this band will be only a factor 5 smaller than pulsations in the optical band.  
Alternatively, an optical mission on the follow-up of the {\sc PLATO} satellite will be able to observe stars only in regions located well above the galactic disc.  In particular, red giant stars could be a very interesting target since these stars can be observed up to distances near the Galactic Center, close to 1000 parsec of the supermassive black hole~\citep{2014ExA....38..249R} located in the Galaxy centre. These stars are known to have acoustic, gravity, and mixed quadrupole modes, all of which can be affected by gravitational radiation. Nevertheless the mode amplitude variations on these stars should be very different from the ones computed for sunlike stars, since their internal structures are very different.
Another possibility on the optical band is to look for pulsating stars in globular clusters and dwarf galaxies. Additionally, the probability of such detections being achieved successfully would increase significantly if the  source of gravitational radiation is located near a population of stars, since in this case the quadrupole modes of several stars are affected simultaneously or contemporaneously within the same field of view.

%\acknowledgments

% % % % % % % % % % % % % % % % % % % % % % % % % % % % % % % % % % % % % % % % % %
\begin{acknowledgments}
The work of I.L. was supported by grants from "Funda\c c\~ao para a Ci\^encia e Tecnologia"  and "Funda\c c\~ao Calouste Gulbenkian". 
Moreover, we are grateful to the authors of the ADIPLS and CESAM codes for having made their codes publicly available.
The author thanks the anonymous referees for the 
insightful	comments and suggestions.
\end{acknowledgments}
%

% % % % % % % % % % % % % % % % % % % % % % % % % % % % % % % % % % % % % % % % % % 
\bibliographystyle{yahapj}

% % % % % % % % % % % % % % % % % % % % % % % % % % % % % % % % % % % % % % % % % % 
%\input{artGw2_biblio}
%\bibliography{libGWart2}
%\bibliography{letGW2bib}
%\include{letgw2_comments}
%\bibliographystyle{yahapj}
%\bibliographystyle{unsrt}
%\bibliographystyle{unsrtnat}
%\bibliography{mnil16lib}
% % % % % % % % % % % % % % % % % % % % % % % % % % % % % % % % % % % % % % % % % %
\end{document}